\documentclass[prd,reprint,showpacs,showkeys]{revtex4}
\usepackage{graphics}
\usepackage{eurosym}
\usepackage{amsfonts}
\usepackage{amssymb}
\usepackage{amsmath}
\usepackage{graphicx}
\usepackage[font={footnotesize,it}]{caption}

\begin{document}

\title{Tunneling of Massive Vector Particles From Rotating Charged Black
Strings}
\author{Kimet Jusufi}
\email{kimet.jusufi@unite.edu.mk}
\author{Ali \"{O}vg\"{u}n}
\email{ali.ovgun@emu.edu.tr}
\affiliation{Physics Department, State University of Tetovo, Ilinden Street nn, 1200,
Macedonia}
\affiliation{Physics Department, Eastern Mediterranean University, Famagusta, Northern
Cyprus, Mersin 10, Turkey}
\date{\today }

\begin{abstract}
We study the quantum tunneling of charged massive vector bosons from a
charged static and a rotating black string. We apply the standard methods,
first we use the WKB approximation and the Hamilton-Jacobi equation, and
then we end up with a set of four linear equations. Finally, solving for the
radial part by using the determinant of the metric equals zero, the
corresponding tunneling rate and the Hawking temperature is recovered in
both cases. The tunneling rate deviates from pure thermality and is
consistent with an underlying unitary theory.
\end{abstract}

\maketitle

\section{Introduction}

In his seminal paper \cite{Hawking1}, Steven Hawking showed that black holes
radiate thermally due to the quantum effects and this radiation is known as
Hawking radiation. Thus, for the first time, it has been established a
relation between thermodynamics and space-time geometry. Furthermore, the
entropy of the black hole is shown to be proportional to the surface area of
the black hole.

Besides the Hawking's original method, today there exists a number of
different approaches deriving the Hawking temperature \cite%
{gibbons1,gibbons2,umetso}. The tunneling method \cite%
{kraus1,kraus2,perkih1,perkih2,perkih3,ang,sri1,sri2,vanzo}, has been
studied in details and shown to be very successful for calculating the
Hawking temperature for different types of particles emitted from static as
well as stationary space-time metrics \cite{sh2,mann0,mann1,mann2,mann3}.
Hawking temperature depends on the black hole mass $M$, charge $Q$ and
angular momentum $J$, using the tunneling approach, it is also shown that,
the Hawking temperature for a particular black hole configuration remains
unaltered and unaffected by the nature of particles emitted from the black
hole. Moreover, the radiation spectrum is shown to deviate from pure
thermality due to the conservation of energy, and hence the theory is
consistent with an underlying unitary theory.

Due to the non-linearity of the Einstein's field equations it is very
difficult to find exact solutions. However, apart from the standard
solutions characterized with spherical symmetry, solutions with cylindrical
symmetry have also been found, such solutions are known as cylindrical black
holes or black strings \cite{lemos,cai}. The tunneling of scalar and Dirac
particles from charged static/rotating black string has been also
investigated \cite{gohar1,gohar2, ahmed1,ahmed2}. Recently, the tunneling of
massive spin-$1$ particles has attracted interest \cite%
{xiang,ali,sh11,sh3,sh4,kruglov1,kruglov2}. Therefore, in this paper, we aim
to study the tunneling of massive vector bosons $W^{\pm}$(spin-$1$
particles) from the space-time of a charged static and a rotating black
string. First, we derive the field equations by using the Lagrangian given
by the Glasgow-Weinberg-Salam model. We then use the WKB approximation and
the separation of variables which results with a set of four linear
equations, solving for the radial part by using the determinant of the
metric equals zero, we found the tunneling rate and the corresponding
Hawking temperature in both cases.

The paper is organized as follows. In Sec. II, we investigate the tunneling
of massive vector particles from the static charged black strings and
calculate the corresponding tunneling rate and the Hawking temperature. In
Sec. III, we extend our calculations for the case of tunneling of massive
vector particles from a rotating charged black string. In Sec. IV, we
comment on our results.

\section{Tunneling From Static Charged Black Strings}

We can begin by writing the Einstein-Hilbert action with a negative
cosmological constant in the presence of an electromagnetic field given by 
\begin{equation}
S=\frac{1}{16\pi G}\int d^{4}x\sqrt{-g}\left(R-2\Lambda\right)-\frac{1}{16
\pi}\int d^{4}x \sqrt{-g}F^{\mu \nu}F_{\mu \nu},
\end{equation}
where the Maxwell electromagnetic tensor is given by 
\begin{equation}
F_{\mu \nu}=\partial_{\mu}A_{\nu}-\partial_{\nu}A_{\mu}.
\end{equation}

If one takes into account the cylindrical symmetries of the space-time, then
the line element for a static charged black string with negative
cosmological constant in the presence of electromagnetic fields is shown to
be \cite{lemos,cai} 
\begin{equation}
ds^{2}=-f(r)dt^{2}+f(r)^{-1}dr^{2}+r^{2}d\theta ^{2}+\alpha^{2}r^{2}dz^{2},
\label{metric1}
\end{equation}%
where 
\begin{equation}
f(r)=\alpha^{2} r^{2}-\frac{b}{\alpha r}+\frac{c^{2}}{\alpha^{2}r^{2}},
\label{f}
\end{equation}
and 
\begin{equation}
\alpha^{2}=-\frac{1}{3}\Lambda,\,\,\,\, b=4GM,\,\,\,\, c^{2}=4GQ^{2}.
\end{equation}

Solving for $\alpha^{2} r^{2}-\frac{b}{\alpha r}+\frac{c^{2}}{\alpha^{2}r^{2}%
}=0$, one can easily find the outer horizon given by \cite{gohar1} 
\begin{equation}
r_{+}=\frac{b^{\frac{1}{3}}\sqrt{s}+\sqrt{2\sqrt{s^{2}-4p^{2}-s}}}{2\alpha},
\end{equation}
where 
\begin{eqnarray}
s&=&\left(\frac{1}{2}+\frac{1}{2}\sqrt{1-4\left(\frac{4p^{2}}{3}\right)^{3}}%
\right)^{\frac{1}{3}}+\left(\frac{1}{2}-\frac{1}{2}\sqrt{1-4\left(\frac{%
4p^{2}}{3}\right)^{3}}\right)^{\frac{1}{3}}, \\
p^{2}&=&\frac{c^{2}}{b^{\frac{4}{3}}}.
\end{eqnarray}

Let us now write the Lagrangian density which describes the $W^{\pm}$-bosons
in a background electromagnetic field given by \cite{xiang} 
\begin{equation}
\mathcal{L}=-\frac{1}{2}\left( D_{\mu }^{+}W_{\nu }^{+}-D_{\nu }^{+}W_{\mu
}^{+}\right) \left( D^{-\mu }W^{-\nu }-D^{-\nu }W^{-\mu }\right) +\frac{%
m_{W}^{2}}{\hbar ^{2}}W_{\mu }^{+}W^{-\mu }-\frac{i}{\hbar }eF^{\mu \nu
}W_{\mu }^{+}W_{\nu }^{-},
\end{equation}%
where $D_{\pm \mu }=\nabla _{\mu }\pm \frac{i}{\hbar }eA_{\mu }$ and $\nabla
_{\mu }$ is the covariant geometric derivative. Also, $e$ gives the charge
of the $W^{+}$ boson, $A_{\mu}$ is the electromagnetic vector potential of
the black string given by $A_{\mu}=(-h(r),0,0,0)$, here $h(r)=2Q/\alpha r$,
where $Q$ is the charge of the black string. Using the above Lagrangian the
equation of motion for the $W$-boson field reads 
\begin{equation}
\frac{1}{\sqrt{-g}}\partial _{\mu }\left[ \sqrt{-g}\left( D^{\pm \nu }W^{\pm
\mu }-D^{\pm \mu }W^{\pm \nu }\right) \right] \pm \frac{ieA_{\mu }}{\hbar }%
\left( D^{\pm \nu }W^{\pm \mu }-D^{\pm \mu }W^{\pm \nu }\right) +\frac{%
m_{W}^{2}}{\hbar ^{2}}W^{\pm \nu }\pm \frac{i}{\hbar }eF^{\mu \nu }W_{\mu
}^{\pm }=0  \label{proca1}
\end{equation}%
where $F^{\mu \nu }=\nabla ^{\mu }A^{\nu }-\nabla ^{\nu }A^{\mu }$. In this
work, we will investigate the tunneling of $W^{+}$ boson, therefore one
needs to solve the following equation 
\begin{equation}
\frac{1}{\sqrt{-g}}\partial _{\mu }\left[ \sqrt{-g}g^{\mu \alpha }g^{\nu
\beta }\left( \partial _{\beta }W_{\alpha }^{+}-\partial _{\alpha }W_{\beta
}^{+}+\frac{i}{\hbar }eA_{\beta }W_{\alpha }^{+}-\frac{i}{\hbar }eA_{\alpha
}W_{\beta }^{+}\right) \right]  \label{proca2}
\end{equation}

\begin{equation*}
+\frac{ieA_{\mu }g^{\mu \alpha }g^{\nu \beta }}{\hbar }\left( \partial
_{\beta }W_{\alpha }^{+}-\partial _{\alpha }W_{\beta }^{+}+\frac{i}{\hbar }%
eA_{\beta }W_{\alpha }^{+}-\frac{i}{\hbar }eA_{\alpha }W_{\beta }^{+}\right)
+\frac{m_{W}^{2}g^{\nu \beta }}{\hbar ^{2}}W_{\beta }^{+}+\frac{i}{\hbar }%
eF^{\nu \alpha }W_{\alpha}^{+}=0,
\end{equation*}%
for $\nu =0,1,2,3$. Using the WKB approximation 
\begin{equation}
W_{\mu }^{+}(t,r,\theta ,z )=C_{\mu }(t,r,\theta,z)\,\exp \left( \frac{i}{\hbar }%
S(t,r,\theta ,z)\right) ,  \label{ans1}
\end{equation}%
where the action is given by 
\begin{equation}
S(t,r,\theta ,z)=S_{0}(t,r,\theta ,z)+\hbar\, S_{1}(t,r,\theta ,z)+\hbar\,
^{2}S_{2}(t,r,\theta ,z)+...  \label{act}
\end{equation}

We can now use the last three equations and neglect the terms of higher
order of $\hbar $, then one can find the following set of four equations: 
\begin{eqnarray}
0 &=&C_{0}\left(-(\partial _{1}S_{0})^{2}-\frac{(\partial _{2}S_{0})^{2}}{%
r^{2}f}-\frac{(\partial _{3}S_{0})^{2}}{\alpha^{2}r^{2}f}-\frac{m^{2}}{f}%
\right) +C_{1}\left((\partial _{1}S_{0})\left( eA_{0}+\partial
_{0}S_{0}\right) \right) +C_{2}\left(\frac{(\partial _{2}S_{0})}{r^{2}f}%
\left( \partial _{0}S_{0}+eA_{0}\right) \right)  \notag \\
&+&C_{3}\left(\frac{(\partial _{3}S_{0})}{\alpha^{2}r^{2}f}\left( \partial
_{0}S_{0}+eA_{0}\right) \right) ,
\end{eqnarray}%
\begin{eqnarray}
0 &=&C_{0}\left( -(\partial _{1}S_{0})(eA_{0}+\partial _{0}S_{0})\right)
+C_{1}\left( -f\frac{(\partial _{2}S_{0})^{2}}{r^{2}}-f\frac{(\partial
_{3}S_{0})^{2}}{\alpha^{2}r^{2} }+(\partial
_{0}S_{0}+eA_{0})^{2}-m^{2}f\right) +C_{2}\left( f\frac{(\partial
_{1}S_{0})(\partial _{2}S_{0})}{r^{2}}\right)  \notag \\
&+&C_{3}\left( f\frac{(\partial _{1}S_{0})(\partial _{3}S_{0})}{%
\alpha^{2}r^{2}}\right) ,
\end{eqnarray}%
\begin{eqnarray}
0 &=&C_{0}\left( -\partial _{2}S_{0}\left( \frac{\partial _{0}S_{0}+eA_{0}}{f%
}\right) \right) +C_{1}\left( f(\partial _{2}S_{0})(\partial
_{1}S_{0})\right) +C_{2}\left( -f(\partial _{1}S_{0})^{2}-\frac{(\partial
_{3}S_{0})^{2}}{\alpha^{2}r^{2}}+\frac{(\partial _{0}S_{0}+eA_{0})^{2}}{f}%
-m^{2}\right)  \notag \\
&+&C_{3}\left( \frac{(\partial _{2}S_{0})(\partial _{3}S_{0})}{%
\alpha^{2}r^{2}}\right) ,
\end{eqnarray}%
\begin{eqnarray}
0 &=&C_{0}\left( -\partial _{3}S_{0}\left( \frac{\partial _{0}S_{0}+eA_{0}}{f%
}\right) \right) +C_{1}\left( f(\partial _{3}S_{0})(\partial
_{1}S_{0})\right) +C_{3}\left( -f(\partial _{1}S_{0})^{2}-\frac{(\partial
_{2}S_{0})^{2}}{r^{2}}+\frac{(\partial _{0}S_{0}+eA_{0})^{2}}{f}-m^{2}\right)
\notag \\
&+&C_{2}\left(\frac{(\partial _{2}S_{0})(\partial _{3}S_{0})}{r^{2}}\right) .
\end{eqnarray}

From the metric \eqref{metric1}, it is clear that due to the space-time
symmetries we can use the following ansatz for the action 
\begin{equation}
S_{0}(t,r,\theta,z)=-Et+W(r)+J_{1}\theta +J_{2}z+C,
\end{equation}%
where $E, J_{1}, J_{2}$ and $C$ are constants. Therefore, the non-zero
elements of the coefficient matrix $\Xi $ are given by 
\begin{eqnarray}
\Xi _{11} &=&-(W^{\prime})^{2}-\frac{J_{1}^{2}}{r^{2}f}-\frac{J_{2}^{2}}{%
\alpha^{2} r^{2}f}-\frac{m^{2}}{f}  \notag \\
\Xi _{12} &=&-\Xi _{21}=W^{\prime}\left(eA_{0}-E\right)  \notag \\
\Xi _{13} &=&\frac{J_{1}}{r^{2}f}\left(eA_{0}-E\right)  \notag \\
\Xi _{14} &=&\frac{J_{2}}{\alpha^{2} r^{2}f}\left(eA_{0}-E\right)  \notag \\
\Xi _{22} &=&\left(-f\frac{J_{1}^{2}}{r^{2}}-f\frac{J_{2}^{2}}{%
\alpha^{2}r^{2}}+(eA_{0}-E)^{2}-m^{2}f\right)  \notag \\
\Xi _{23} &=&f\frac{W^{\prime}J_{1}}{r^{2}}  \notag \\
\Xi _{24} &=&f\frac{W^{\prime}J_{2}}{\alpha^{2} r^{2}}  \notag \\
\Xi _{31} &=&-J_{1}\frac{\left(eA_{0}-E\right)}{f}  \notag
\end{eqnarray}
\begin{eqnarray}
\Xi _{32} &=&f J_{1}W^{\prime}  \notag \\
\Xi _{33} &=&\left( -f(W^{\prime})^{2}-\frac{J_{2}^{2}}{\alpha^{2} r^{2}}+%
\frac{(eA_{0}-E)^{2}}{f}-m^{2}\right)  \notag \\
\Xi _{34} &=&\frac{J_{1}J_{2}}{\alpha^{2} r^{2}}  \notag \\
\Xi _{41} &=&-\frac{J_{2}\left(eA_{0}-E\right) }{f}  \notag \\
\Xi _{42} &=&fJ_{2} W^{\prime}  \notag \\
\Xi _{43}&=&\frac{J_{1}J_{2}}{r^{2}}  \notag \\
\Xi _{44} &=&\left( -f(W^{\prime})^{2}-\frac{J_{1}^{2}}{r^{2}}+\frac{%
(eA_{0}-E)^{2}}{f}-m^{2}\right) .
\end{eqnarray}

The nontrivial solution of this equation \cite{kruglov1} 
\begin{equation}
\Xi (C_{0},C_{1},C_{2},C_{3})^{T}=0,  \label{matrixeq}
\end{equation}
is obtained by using the determinant of the matrix equals zero, $\det \Xi=0 $%
, it follows 
\begin{equation}
m^{2}\Big(-r^{2}(E-eA_{0})^{2}\alpha^{2}+f^{2}r^{2}\alpha^{2}(W^{%
\prime})^{2}+\left((m^{2}r^{2}+J_{1}^{2})\,\alpha^{2}+J_{2}^{2}\right)f\Big)^{3}=0.
\end{equation}

Solving this equation for the radial part leads to the following integral 
\begin{equation}
W_{\pm }(r)=\pm \int \frac{\sqrt{(E-eA_{0})^{2}-f(r)\left( m^{2}+\frac{%
J_{1}^{2}}{r^{2}}+\frac{J_{2}^{2}}{\alpha^{2}r^{2}}\right)}}{f(r)}dr.
\end{equation}

Expanding the function $f(r)$ in Taylor's series near the horizon 
\begin{equation}
f(r_{+})\approx f^{\prime }(r_{+})(r-r_{+}),
\end{equation}%
and by integrating around the pole at the outer horizon $r_{+}$, gives 
\begin{equation}
W_{\pm }(r)=\pm \frac{i\pi (E-eA_{0})}{f^{\prime }(r_{+})}.  \label{integral}
\end{equation}

Now we can set the probability of the ingoing particle to $100\%$ (since
every outside particle falls into the black hole), it follows 
\begin{equation*}
P_{-}\simeq e^{-2ImW_{-}}=1,
\end{equation*}%
which implies $ImC=-ImW_{-}$. For the outgoing particle we have $%
ImS_{+}=ImW_{+}+ImC$, and also we make use of $W_{+}=-W_{-}$, which leads to
the probability for the outgoing particle given by 
\begin{equation}
P_{+}=e^{-2ImS}\simeq e^{-4ImW_{+}}.
\end{equation}

In this way the tunneling rate of particles tunneling from inside to outside
the horizon is given by 
\begin{equation}
\Gamma =\frac{P_{+}}{P_{-}}\simeq e^{(-4ImW_{+})}.
\end{equation}

We can find the Hawking temperature simply by compering the last result with
the Boltzmann factor $\Gamma =e^{-\beta E_{net}}$, where $E_{net}=(E-eA_{0})$
and $\beta =1/T_{H}$, yielding 
\begin{equation}
T_{H}=\frac{f^{\prime }(r_{+})}{4\pi }.
\end{equation}

Using Eqn.\eqref{f}, one can recover the Hawking temperature for a static
charged black string \cite{gohar1} 
\begin{equation}
T_{H}=\frac{1}{4\pi}\left(2\alpha^{2}r_{+}+\frac{b}{\alpha r_{+}^{2}}-\frac{%
2c^{2}}{\alpha^{2}r_{+}^{3}}\right).
\end{equation}

\section{Tunneling From Rotating Charged Black Strings (RCBSs)}

Lemos derived a rotating charged cylindrically symmetric exact solution of
Einstein equations for a black string \cite{lemos}. The line element for a
RCBSs is given by \cite{gohar1}

\begin{equation}
ds^{2}=-F(r)\,dt^{2}+R^{2}(r)\left(N dt+d\theta \right) ^{2}+\frac{dr^{2}}{%
G(r)}+\alpha ^{2}r^{2}dz^{2},
\end{equation}

where the lapse function $F$ and the shift function $N$ are given as

\begin{equation}
G=\left( \alpha ^{2}r^{2}-\frac{b}{\alpha r}+\frac{c^{2}}{\alpha ^{2}r^{2}}%
\right) ,  \label{G}
\end{equation}%
\begin{equation}
F=fG,
\end{equation}%
\begin{equation}
f=\left( \gamma ^{2}-\frac{\omega ^{2}}{\alpha ^{2}}\right) ^{2}\frac{r^{2}}{%
R^{2}},
\end{equation}

\begin{equation}
N=-\frac{\gamma \omega }{\alpha ^{2}R^{2}}\left( \frac{b}{\alpha r}-\frac{%
c^{2}}{\alpha ^{2}r^{2}}\right) ,
\end{equation}%
and 
\begin{equation}
R^{2}=\gamma ^{2}r^{2}-\frac{\omega ^{2}}{\alpha ^{4}}\left( \alpha
^{2}r^{2}-\frac{b}{\alpha r}+\frac{c^{2}}{\alpha ^{2}r^{2}}\right).
\end{equation}

Noted that the rotation parameter $a=J/M,$ constant $\alpha ^{2}=-\Lambda /3$%
, where $\Lambda $ is the cosmological constant, $M$ is the \ ADM mass, $Q$
is the charge of the black string, and $J$ is the angular momentum. In
addition, $b$ and $c$ are defined as%
\begin{equation}
b=4M\left( 1-\frac{3a^{2}\alpha ^{2}}{2}\right) ,
\end{equation}%
\begin{equation}
c^{2}=4Q^{2}\left( \frac{1-3a^{2}\alpha ^{2}/2}{1-a^{2}\alpha ^{2}/2}\right)
.
\end{equation}%
\newline
Furthermore, $\gamma ^{2}$ and $\omega ^{2}/\alpha ^{2}$ are defined as 
\begin{equation}
\gamma ^{2}=\frac{2GM}{b}\pm \frac{2G}{b}\sqrt{M^{2}-\frac{8J\alpha ^{2}}{9}}%
,
\end{equation}%
\begin{equation}
\frac{\omega ^{2}}{\alpha ^{2}}=\frac{4GM}{b}\mp \frac{4G}{b}\sqrt{M^{2}-%
\frac{8J\alpha ^{2}}{9}},
\end{equation}%
or 
\begin{equation*}
\gamma =\sqrt{\frac{1-\frac{a^{2}\alpha ^{2}}{2}}{1-\frac{3a^{2}\alpha ^{2}}{%
2}}},
\end{equation*}%
\begin{equation}
\omega =\frac{a\alpha ^{2}}{\sqrt{1-\frac{3a^{2}\alpha ^{2}}{2}}}.
\label{2aaa}
\end{equation}

Let us now introduce the electromagnetic field associated with the vector
potential of the RCBSs 
\begin{equation}
A_{\mu }=(A_{0},0,A_{2},0)
\end{equation}%
where $A_{0}=-\gamma h(r)$, $A_{2}=$ $\frac{\omega }{\alpha ^{2}}h(r),$ and $%
h(r)$ is an arbitrary function of $r$ for the line charge density along the $%
z$-line given by $Q=\frac{Q_{z}}{\Delta z}=\gamma \lambda $. To exactly
reveal the massive vector particle's tunneling radiation, we should solve
the Proca equation in Eqn.(\ref{proca2}). Following the standard procedure,
we use the WKB approximation Eqn.(\ref{ans1}) with the action Eqn(\ref{act})
in the background of the RCBSs spacetime and neglect the factors of higher
orders of $\hbar $. Then using the following ansatz for the action 
\begin{equation}
S_{0}=-Et+W(r)+J_{1}\theta +J_{2}z+k,
\end{equation}%
where $E,J_{1},J_{2}$ and $k$ are constants, we get four decoupled equations
such as:

\begin{equation*}
\frac{C_{0}}{fG^{2}R^{2}r^{2}\alpha ^{2}}\Big[fG^{3}R^{2}r^{2}\alpha
^{2}W^{\prime 2}+G\Big[\left( \left( m^{2}r^{2}\alpha ^{2}+J_{2}^{2}\right)
fG-r^{2}\left( eA_{2}+J_{1}\right) N\alpha ^{2}\left( \left(
eA_{2}+J_{1}\right) N-eA_{0}+E\right) \right) R^{2}
\end{equation*}%
\begin{equation*}
+r^{2}\alpha ^{2}fG\left( eA_{2}+J_{1}\right) ^{2}\Big]\Big]-\left( \left(
eA_{2}+J_{1}\right) N-eA_{0}+E\right) \frac{W^{\prime }}{fG}C_{1}+\frac{%
\left( -erA_{2}-J_{1}r\right) }{fGr}\left( \left( eA_{2}+J_{1}\right)
N-eA_{0}+E\right) C_{2}
\end{equation*}%
\begin{equation}
-\frac{C_{3}J_{2}}{fG}\left( \left( eA_{2}+J_{1}\right) N-eA_{0}+E\right) =0,
\end{equation}

\begin{equation*}
\frac{C_{0}}{fGR^{2}\alpha ^{2}r^{2}}\left( -\alpha
^{2}fG^{2}R^{2}eA_{0}r^{2}W^{\prime }+\alpha ^{2}R^{2}fG^{2}W^{\prime
}r^{2}E\right) +2\Big[\alpha ^{2}\left( \left( \left( -N^{2}A_{2}+NA_{0}\right)
R^{2}+fGA_{2}\right) e-R^{2}NE\right) r^{2}J_{1}
\end{equation*}%
\begin{equation*}
+\frac{1}{2}\alpha ^{2}r^{2}\left( -R^{2}N^{2}+fG\right) J_{1}^{2}-\alpha
^{2}R^{2}e\left( NA_{2}-A_{0}\right) r^{2}E
\end{equation*}%
\begin{equation*}
-\frac{1}{2}R^{2}\alpha ^{2}r^{2}E^{2}+\alpha ^{2}r^{2}\left( \frac{1}{2}%
\left( m^{2}fG-e^{2}\left( NA_{2}-A_{0}\right) ^{2}\right) R^{2}+\frac{1}{2}%
fGe^{2}A_{2}^{2}\right) +\frac{1}{2}R^{2}fGJ_{2}^{2}\Big]\frac{C_{1}}{%
fGR^{2}\alpha ^{2}r^{2}}
\end{equation*}%
\begin{equation}
+\Big[ -\alpha ^{2}fG^{2}R^{2}A_{2}er^{2}W^{\prime }-\alpha
^{2}G^{2}fR^{2}W^{\prime }r^{2}J_{1}\Big] \frac{C_{2}}{fGR^{2}\alpha
^{2}r^{2}}-GJ_{2}W^{\prime }C_{3}=0,
\end{equation}

\begin{equation*}
\Big[-\alpha ^{2}\left( -fG^{2}R\left( -R^{2}N^{2}+fG\right)
rE-fG^{2}\left( N^{2}A_{0}erR^{2}-fGA_{0}er\right) R\right)
rJ_{1}-\alpha^{2}RfG^{2}\left( \left( reA_{2}N^{2}-2A_{0}erN\right)
R^{2}-fGA_{2}er\right)rE
\end{equation*}
\begin{equation*}
-\alpha ^{2}r^{2}fG^{2}R^{3}NE^{2}+\alpha ^{2}fG^{2}\left( e^{2}\left(
NA_{2}-A_{0}\right) A_{0}rNR^{2}-fGA_{2}A_{0}e^{2}r\right) Rr\Big]\frac{C_{0}}{%
f^{2}G^{3}R^{3}r^{2}\alpha ^{2}}+\Big[-\alpha
^{2}r^{2}fG^{2}R\left(-R^{2}N^{2}+fG\right) W^{\prime }J_{1}
\end{equation*}%
\begin{equation*}
-2\alpha ^{2}fG^{2}R\left( -\frac{1}{2}erN\left( NA_{2}-A_{0}\right) R^{2}+%
\frac{1}{2}fGA_{2}er\right) rW^{\prime }+\alpha
^{2}R^{3}fG^{2}Nr^{2}W^{\prime }E\Big]\frac{C_{1}}{f^{2}G^{3}R^{3}r^{2}\alpha
^{2}}
\end{equation*}%
\begin{equation*}
+\Big[ -\alpha ^{2}\left( fG^{2}R^{3}NrE-fG^{2}NA_{0}erR^{3}\right)
rJ_{1}+f^{2}G^{3}R^{3}J_{2}^{2}+\alpha ^{2}fG^{2}\left( fGrm^{2}+e^{2}\left(
NA_{2}-A_{0}\right) A_{0}r\right) R^{3}r
\end{equation*}%
\begin{equation*}
-\alpha ^{2}R^{3}fG^{2}\left( NA_{2}er-2A_{0}er\right)
rE-\alpha^{2}r^{2}fG^{2}R^{3}E^{2}+r^{2}\alpha ^{2}R^{3}f^{2}G^{4}W^{\prime
2}\Big]\frac{C_{2}}{f^{2}G^{3}R^{3}r^{2}\alpha ^{2}}
\end{equation*}

\begin{equation}
+\Big[-\alpha ^{2}r^{2}fG^{2}R\left( -R^{2}N^{2}+fG\right) J_{2}J_{1}+\alpha
^{2}R^{3}r^{2}fG^{2}NJ_{2}E-\alpha ^{2}fG^{2}R\left( \left(
-N^{2}A_{2}+NA_{0}\right) R^{2}+fGA_{2}\right) er^{2}J_{2}\Big]C_{3}=0,
\end{equation}

\begin{equation}
\left( -fG^{2}A_{0}erR+RrG^{2}fE\right) \frac{J_{2}}{R\alpha ^{2}r^{3}fG^{2}}%
C_{0}-\frac{W^{\prime }J_{2}C_{1}}{r^{2}\alpha ^{2}}+\left(
-RfG^{2}A_{3}er-J_{1}RfG^{2}r\right) \frac{J_{2}}{R\alpha ^{2}r^{3}fG^{2}}%
C_{2}
\end{equation}%
\begin{equation*}
-\Big[r\left( -R^{2}N^{2}+fG\right) J_{1}^{2}-2\left( \left( \left(
-N^{2}A_{2}+NA_{0}\right) R^{2}+fGA_{2}\right) e-R^{2}NE\right)
rJ_{1}-rR^{2}fG^{2}W^{\prime 2}+rR^{2}E^{2}
\end{equation*}%
\begin{equation*}
+2erR^{2}\left( NA_{2}-A_{0}\right) E-\left( \left( m^{2}fG-e^{2}\left(
NA_{2}-A_{0}\right) ^{2}\right) R^{2}+fGe^{2}A_{2}^{2}\right) r\Big]\frac{C_{3}}{%
GrfR^{2}}=0.
\end{equation*}

Then the non-zero elements of the coefficient matrix $\Theta $ are
calculated as following%
\begin{eqnarray}
\Theta _{11} &=&\Big[fG^{3}R^{2}r^{2}\alpha ^{2}W^{\prime 2}+G\Big[\left( \left(
m^{2}r^{2}\alpha ^{2}+J_{2}^{2}\right) fG-r^{2}\left( eA_{2}+J_{1}\right)
N\alpha ^{2}\left( \left( eA_{2}+J_{1}\right) N-eA_{0}+E\right) \right) R^{2}
\\
&&+r^{2}\alpha ^{2}fG\left( eA_{2}+J_{1}\right) ^{2}\Big]\Big],  \notag \\
\Theta _{12} &=&-\left( \left( eA_{2}+J_{1}\right) N-eA_{0}+E\right)
W^{\prime },  \notag \\
\Theta _{13} &=&\left( -erA_{2}-J_{1}r\right) \left( \left(
eA_{2}+J_{1}\right) N-eA_{0}+E\right) ,  \notag \\
\Theta _{14} &=&-J_{2}\left( \left( eA_{2}+J_{1}\right) N-eA_{0}+E\right) , 
\notag \\
\Theta _{21} &=&\left( -\alpha ^{2}fG^{2}R^{2}eA_{0}r^{2}W^{\prime }+\alpha
^{2}R^{2}fG^{2}W^{\prime }r^{2}E\right) ,  \notag \\
\Theta _{22} &=&2\Big[\alpha ^{2}\left( \left( \left( -N^{2}A_{2}+NA_{0}\right)
R^{2}+fGA_{2}\right) e-R^{2}NE\right) r^{2}J_{1}+\frac{1}{2}\alpha
^{2}r^{2}\left( -R^{2}N^{2}+fG\right) J_{1}^{2}-\alpha ^{2}R^{2}e\left(
NA_{2}-A_{0}\right) r^{2}E  \notag \\
&&-\alpha ^{2}r^{2}fG^{2}R^{3}NE^{2}+\alpha ^{2}r^{2}\left( \frac{1}{2}%
\left( m^{2}fG-e^{2}\left( NA_{2}-A_{0}\right) ^{2}\right) R^{2}+\frac{1}{2}%
fGe^{2}A_{2}^{2}\right) +\frac{1}{2}R^{2}fGJ_{2}^{2}\Big],  \notag \\
\Theta _{23} &=&\left[ -\alpha ^{2}fG^{2}R^{2}A_{2}er^{2}W^{\prime }-\alpha
^{2}G^{2}fR^{2}W^{\prime }r^{2}J_{1}\right] ,  \notag \\
\Theta _{24} &=&-GJ_{2}W^{\prime },  \notag \\
\Theta _{31} &=&\Big[-\alpha ^{2}\left( -fG^{2}R\left( -R^{2}N^{2}+fG\right)
rE-fG^{2}\left( N^{2}A_{0}erR^{2}-fGA_{0}er\right) R\right) rJ_{1}  \notag \\
&&-\alpha ^{2}RfG^{2}\left( \left( reA_{2}N^{2}-2A_{0}erN\right)
R^{2}-fGA_{2}er\right) rE-\alpha ^{2}r^{2}fG^{2}R^{3}NE^{2}+\alpha ^{2}fG^{2}
\notag \\
&&+\left( e^{2}\left( NA_{2}-A_{0}\right)
A_{0}rNR^{2}-fGA_{2}A_{0}e^{2}r\right) Rr\Big],  \notag \\
\Theta _{32} &=&\Big[-\alpha ^{2}r^{2}fG^{2}R\left( -R^{2}N^{2}+fG\right)
W^{\prime }J_{1}-2\alpha ^{2}fG^{2}R\left( -\frac{1}{2}erN\left(
NA_{2}-A_{0}\right) R^{2}+\frac{1}{2}fGA_{2}er\right) rW^{\prime }  \notag \\
&&+\alpha ^{2}R^{3}fG^{2}Nr^{2}W^{\prime }E\Big],  \notag \\
\Theta _{33} &=&\Big[-\alpha ^{2}\left(
fG^{2}R^{3}NrE-fG^{2}NA_{0}erR^{3}\right)
rJ_{1}+f^{2}G^{3}R^{3}J_{2}^{2}+\alpha ^{2}fG^{2}\left( fGrm^{2}+e^{2}\left(
NA_{2}-A_{0}\right) A_{0}r\right) R^{3}r  \notag \\
&&-\alpha ^{2}R^{3}fG^{2}\left( NA_{2}er-2A_{0}er\right) rE-\alpha
^{2}r^{2}fG^{2}R^{3}E^{2}+r^{2}\alpha ^{2}R^{3}f^{2}G^{4}W^{\prime 2}\Big], 
\notag \\
\Theta _{34} &=&\Big[-\alpha ^{2}r^{2}fG^{2}R\left( -R^{2}N^{2}+fG\right)
J_{2}J_{1}+\alpha ^{2}R^{3}r^{2}fG^{2}NJ_{2}E-\alpha ^{2}fG^{2}R\left(
\left( -N^{2}A_{2}+NA_{0}\right) R^{2}+fGA_{2}\right) er^{2}J_{2}\Big],  \notag
\\
\Theta _{41} &=&\left( -fG^{2}A_{0}erR+RrG^{2}fE\right) J_{2},  \notag \\
\Theta _{42} &=&-W^{\prime }J_{2},  \notag \\
\Theta _{43} &=&\left( -RfG^{2}A_{2}er-J_{1}RfG^{2}r\right) J_{2},  \notag \\
\Theta _{44} &=&-\Big[r\left( -R^{2}N^{2}+fG\right) J_{1}^{2}-2\left( \left(
\left( -N^{2}A_{2}+NA_{0}\right) R^{2}+fGA_{2}\right) e-R^{2}NE\right)
rJ_{1}-rR^{2}fG^{2}W^{\prime 2}  \notag \\
&&+rR^{2}E^{2}+2erR^{2}\left( NA_{2}-A_{0}\right) E-\left( \left(
m^{2}fG-e^{2}\left( NA_{2}-A_{0}\right) ^{2}\right)
R^{2}+fGe^{2}A_{2}^{2}\right) r\Big].  \notag
\end{eqnarray}

The nontrivial solution of this equation \cite{kruglov1} 
\begin{equation}
\Theta (C_{0},C_{1},C_{2},C_{3})^{T}=0,  \label{matrixeq}
\end{equation}%
is obtained by using the determinant of the matrix equals zero, $\det \Theta
=0$, it follows 
\begin{equation}
-m^{2}\left[ -fG^{2}R^{2}r^{2}\alpha ^{2}W^{\prime 2}+\left( -f\left(
m^{2}r^{2}\alpha ^{2}+J_{2}^{2}\right) G+r^{2}\alpha ^{2}\left( \left(
eA_{2}+J_{1}\right) N-eA_{0}+E\right) ^{2}\right) R^{2}-Gf\alpha
^{2}r^{2}\left( eA_{2}+J_{1}\right) ^{2}\right] ^{3}=0.
\end{equation}

Solving this equation for the radial part leads to the following integral,
as noted that $F(r)=f(r)G(r),$ 
\begin{equation}
W_{\pm }(r)=\pm \int \frac{R(r)\sqrt{\left( E-eA_{0}+\left(
eA_{2}+J_{1}\right) N\right) ^{2}-F\left[\left( m^{2}+\frac{J_{2}^{2}}{%
r^{2}\alpha ^{2}}\right)+\frac{\left( eA_{2}+J_{1}\right) ^{2}}{R^{2}}\right]%
}}{\left(\gamma^{2}-\frac{\omega^{2}}{\alpha^{2}}\right)r \,G(r)}dr.
\end{equation}

Integrating around the pole at the outer horizon $r_{+}$, and by using $%
R(r_{+})=\gamma r_{+}$, gives \cite{ang,sri1} 
\begin{equation}
W_{\pm }(r)=\pm \frac{i\pi \gamma \left( E-eA_{0}+\left( eA_{2}+J_{1}\right)
N\right) }{\left(\gamma^{2}-\frac{\omega^{2}}{\alpha^{2}}\right)G^{\prime
}(r_{+})},
\end{equation}
where $E_{net}=\left( E-eA_{0}+\left( eA_{2}+J_{1}\right) N\right).$ By the
same way used in the first part, the tunneling rate of particles tunneling
from inside to outside the horizon is given by 
\begin{equation}
\Gamma =\frac{P_{+}}{P_{-}}\simeq e^{(-4ImW_{+})}.
\end{equation}

On the other hand, using Eqns.\eqref{G} and \eqref{2aaa}, it follows 
\begin{equation}
\gamma^{2}-\frac{\omega^{2}}{\alpha^{2}}=1,
\end{equation}
and 
\begin{equation}
G^{\prime}(r_{+})=\left(2\alpha^{2}r_{+}+\frac{b}{\alpha r_{+}^{2}}-\frac{%
2c^{2}}{\alpha^{2}r_{+}^{2}}\right).
\end{equation}

Again, comparing the Boltzmann factor $\Gamma =e^{-\beta E_{net}}$, with the
tunneling rate, gives the Hawking temperature \cite{gohar2,ahmed1}

\begin{equation}
T_{H}=\frac{G^{\prime }(r_{+})}{4\pi }\frac{\left( \gamma ^{2}-\frac{\omega
^{2}}{\alpha ^{2}}\right) }{\gamma }=\frac{1}{4\pi \gamma }\left( 2\alpha
^{2}r_{+}+\frac{b}{\alpha r_{+}^{2}}-\frac{2c^{2}}{\alpha ^{2}r_{+}^{2}}%
\right) .
\end{equation}

\section{Conclusion}

To summarize, in this paper, we \ derive the charged black strings
temperature using the Hamilton-Jacobi method of the tunneling formalism for
the massive vector particles. In the case of a static black string, we start
from the field equations, then we use the WKB approximation and the
separation of variables which results with a set of four equations. In order
to work out the Hawking temperature, we solve the radial part by using the
determinant of the metric equals zero. Next, we extend our results to the
rotating case and calculate the Hawking temperature. Finally, the results
presented in this work extend the tunneling method for massive \ vector
bosons in the case of static/rotating black strings and are consistent with
those in the literature\cite{gohar1,gohar2,ahmed1,ahmed2}.

\section{ACKNOWLEDGMENT}

The authors would like to thank the editor and the anonymous reviewers.

\end{document}